\documentclass[aps, prd, twocolumn, lengthcheck, superscriptaddress, nofootinbib]{revtex4-2} 
\usepackage{amsfonts}
\usepackage{amsmath}
\usepackage{amssymb}
\usepackage{xcolor}
\usepackage{graphicx}
\usepackage{subfigure}
\usepackage{slashed}
\usepackage{caption}
\usepackage{hyperref}
\usepackage{amsthm}
\usepackage{mathrsfs}
\usepackage{color}
\usepackage{epsfig}
\usepackage{epstopdf}
\usepackage{bm}
\usepackage[T1]{fontenc}
\usepackage{ae,aecompl}	
\usepackage[toc,page]{appendix}
\usepackage{soul}
\usepackage[utf8]{inputenc}
\usepackage{multirow}
\usepackage{verbatim}
\usepackage{natbib}

\def\be{\begin{eqnarray}}\def\ee{\end{eqnarray}}

\begin{document}
	\bibliographystyle{plain}

\title{White Dwarf Structure and Binary Inspiral Gravitational Waves from Quantum Hadrodynamics}

\author{Ling-Jun Guo}

\affiliation{School of Fundamental Physics and Mathematical Sciences, Hangzhou Institute for Advanced Study, UCAS, Hangzhou, 310024, China}
\affiliation{College of Physics, Jilin University, Changchun, 130012, China}

\author{Yao Ma}
\email{mayao@ucas.ac.cn}
\affiliation{School of Fundamental Physics and Mathematical Sciences, Hangzhou Institute for Advanced Study, UCAS, Hangzhou, 310024, China}
\affiliation{School of Frontier Sciences, Nanjing University, Suzhou 215163, China}
	
\author{Yong-Liang Ma}
\email{ylma@nju.edu.cn}
\affiliation{School of Frontier Sciences, Nanjing University, Suzhou 215163, China}
\affiliation{International Center for Theoretical Physics Asia-Pacific (ICTP-AP) , UCAS, Beijing, 100190, China}

\author{Ruo-Xi Wu}
\affiliation{School of Fundamental Physics and Mathematical Sciences, Hangzhou Institute for Advanced Study, UCAS, Hangzhou, 310024, China}
\affiliation{Institute of Theoretical Physics, Chinese Academy of Sciences, Beijing, 100190, China}
\affiliation{University of Chinese Academy of Sciences, Beiing, 100049, China}

\author{Yue-Liang Wu}
\email{ylwu@ucas.ac.cn}
\affiliation{School of Fundamental Physics and Mathematical Sciences, Hangzhou Institute for Advanced Study, UCAS, Hangzhou, 310024, China}
\affiliation{International Center for Theoretical Physics Asia-Pacific (ICTP-AP) , UCAS, Beijing, 100190, China}
\affiliation{Institute of Theoretical Physics, Chinese Academy of Sciences, Beijing, 100190, China}
\affiliation{TaiJi Laboratory for Gravitational Wave Universe (Beijing/Hangzhou), University of Chinese Academy of Sciences, Beijing, 100049, China}

\date{\today}

\begin{abstract}

White dwarfs, one of the compact objects in the universe, play a crucial role in astrophysical research and provide a platform for exploring nuclear physics. In this work, we extend the relativistic mean field approach by using a Walecka-type quantum hadrodynamics model to capture the intricate structure of white dwarfs. We calculate nuclear properties, Coulomb energy, and photon energy within white dwarfs in a unified framework. By carefully calibrating the model parameters to align with nuclear matter properties, we successfully reproduce the structures of several elements in white dwarfs, such as the isotopes of $\rm C$ and $^{16}\rm O$, except for the unnaturally deeply bound state $^4$He. Furthermore, we predict the characteristics of white dwarfs composed of atom-like units and the gravitational waves stemming from binary white dwarf inspirals incorporating tidal deformability contributions up to the 2.5 post-Newtonian order. These results shed light on the structure of white dwarfs and provide valuable information for future gravitational wave detection. This methodological advancement allows for a cohesive analysis of white dwarfs, neutron stars, and the nuclear pasta within a unified theoretical framework.

\end{abstract}

\maketitle

\allowdisplaybreaks

\section{Introduction}

The detection of gravitational waves (GWs) by the LIGO/Virgo Collaborations~\cite{LIGOScientific:2016aoc,LIGOScientific:2017vwq,LIGOScientific:2018mvr} signifies that astronomy has entered the era of multi-messenger observations. This advancement provids a new means to investigate the nature of compact objects such as black holes (BHs) and neutron stars (NSs). The space-based detectors like LISA~\cite{LISA:2017pwj}, Taiji~\cite{Hu:2017mde}, and Tianqin~\cite{TianQin:2015yph} which are scheduled to be launched in the 2030s aim to the GWs with frequency $0.1$mHz-$0.1$Hz. One of the sources of these space-based low-frequency GW detectors is the white dwarf (WD) binaries~\cite{Ruiter:2007xx,Yu:2010fq,McNeill:2019rct,Wolz:2020sqh,Korol:2020lpq,Gibney:2024nature}---another kind of compact object in the universe.

As one of the compact objects in the universe, WDs are stabilized by the equilibrium between the pressure of the degenerate electron gas and inner gravity. Due to the charge neutrality, in addition to electrons, there should be nuclei in WDs. In the pioneer work by Chandrasekhar~\cite{Chandrasekhar:1931ih}, the WD matter was assumed to be the uniformed degenerate electron gas and pointlike nuclei. Then, by introducing the concept of Wigner-Seitz cell, Salpeter studied the corrections due to the non-uniformity of the electrons inside a Wigner-Seitz cell and found that the Coulomb force is essential for determining the maximum stable mass of non-rotating WDs~\cite{Salpeter:1961zz}. In Refs.~\cite{Rotondo:2009cr,Rotondo:2011zz}, a relativistic Feynman-Metropolis-Teller model was considered to include the finite size of nuclei by assuming a constant distribution of protons confined in a radius characterized by the pion Compton wavelength. This model allows for a more accurate calculation of the energy and pressure of the Wigner-Seitz and consequently a more accurate equation of state (EOS) of WD matter. In the WD matter, the pressure is dominated by the degenerate electron gas, while the energy of a Wigner-Seitz cell is mainly from nucleus, and its binding energy is \(>10\rm MeV\) which is already larger than the contribution from electromagnetical dynamics.
Therefore, a realistic description of the nucleus, radius and binding energy, requiring an accurate understanding of the nuclear force which is crucial for the study of WDs.

Binary WDs are closely connected to Type Ia supernova (SN) explosions, which are events rich in physics, involving strong interaction, weak interaction, and intense electromagnetic fields~\cite{Niemeyer:1996pm,Fryer:1998jb,Rosswog:2009fh,Toonen:2012jj}. Understanding the structure of WDs, especially the components of the cores of WDs, is a complex yet crucial task for nuclear physics.  GWs stemming from binary WD systems provide valuable insights into the characteristics of the internal structure of WDs and therefore the nuclear forces~\cite{McNeill:2019rct,Saumon:2022gtu}.

In the description of nuclear force, quantum hadrodynamics (QHD) written in terms of colorless hadrons plays an indispensable role~\cite{Walecka:1974qa,Serot:1984ey,Weinberg:1990rz,Serot:1997xg}. In the realm of hadron interactions, linear and non-linear realizations of chiral symmetry both are frameworks that more directly mirror QCD  properties~\cite{Weinberg:1990rz,Weinberg:1991um,Gasser:1987rb}, providing an alternative to the traditional Walecka-type parametrization. With respect to the flavor or chiral symmetry, the electromagnetic and weak forces can be self-consistently included. Therefore, QHD serves as a unified framework for nucleons, nuclei, and their electromagnetic and weak processes.

In this work, anchored on the Walecka-type QHD including hadrons, electrons and photons, we extend the relativistic mean field (RMF) approach which has been widely used in the study of homogeneous nuclear matter (NM) and NM clusters~\cite{Xia:2022rfc, Xia:2022dvw} to study WD properties. The WDs are suggested to be made of charge-neutral objects---atom-like units---which have finite nucleon number nuclei surrounded by an electron cloud. These objects effectively replicate the Wigner-Seitz cell structure of WDs~\cite{Salpeter:1961zz,Rotondo:2011zz} but with the extension of the nucleus calculated from the Walecka-type model. The approach developed here can be taken as a microscopic materialization of the phenomenological parametrization of EOS of WD matter and a unified framework for NSs, NS crusts, and WDs.
Note that we will not consider the strange degrees of freedom which are found to shrink the size of white dwarfs~\cite{Kurban:2020xtb} since it is beyond the purpose this work.

Using the aforementioned extension, we calculate the properties of $\rm ^{4}He$, isotopes of $\rm C$ and $\rm ^{16}O$ with model parameters carefully fitted by the properties of NM and the structures of NSs. We then derive the structures of WDs, including their mass-radius (M-R) relations and tidal deformability (TD), for various compositions (\(\rm ^{4}He\), \(\rm ^{12}C\), and \(\rm ^{16}O\)). Finally, we study the GW signals stemming from the inspirals of the binary WDs with and without TD through the application of the post-Newtonian (PN) approximation~\cite{Cutler:1992tc,Damour:2000zb} at the 2.5PN order. Our results of WDs made of $\rm ^{4}He$, isotopes of $\rm C$ and $\rm ^{16}O$ are consistent with the observations and Chandrasekhar limit~\cite{Chandrasekhar:1931ih}, which assumes a free electron gas. We also investigate the GW signals stemming from the binary WDs composed of different elements. These results may serve as preliminary guides for future GW detection, providing insights into the intricacies of binary WD dynamics.

The rest of this paper is organized as follows: Sec.~\ref{sec:rmf} introduces the framework of the extended RMF method and presents numerical results pertaining to the properties of NM and nuclei. Sec.~\ref{sec:wd} explores the structures of WDs and simulates the GW signals resulting from binary WD inspirals. Our discussion and outlook are given in the last section. We describe the structures of nuclei and the EOS of WDs in Appendix A and the detail of PN expansion in Appendix B.

\section{The extended RMF and nucleus}
\label{sec:rmf}

Without loss of generality, we consider a Walecka-type model which has been widely used in nuclear physics~\cite{Sugahara:1993wz,Shen:1998gq}. Including electromagnetic (EM) interaction, the model is written as:
\be
\label{eq:lag}
\mathcal{L} & = & \mathcal{L}_{\rm fermion} + \mathcal{L}_{\rm boson} +\mathcal{L}_I\ ,
\ee
where
\be
\mathcal{L}_{\rm fermion} & = & \bar{\Psi}\left(i \slashed{\text D} - m_N\right)\Psi + \bar{\psi}\left(i \slashed{\text D} - m_e\right)\psi\ , \nonumber\\
\mathcal{L}_{\rm boson} & = & \frac{1}{2}(\partial_{\mu}\sigma\partial^{\mu}\sigma-m_{\sigma}^2\sigma^2)-\frac{1}{3}g_2\sigma^3-\frac{1}{4}g_3\sigma^4 \nonumber\\
& &{} - \frac{1}{4}\Omega_{\mu\nu}\Omega^{\mu\nu} + \frac{1}{2}m_{\omega}^2\omega_{\mu}\omega^{\mu}+\frac{1}{4}c_3\left(\omega_{\mu}\omega^{\mu}\right)^2 \nonumber\\
& &{} -\frac{1}{4}\vec{\bm P}_{\mu\nu}\cdot\vec{\bm P}^{\mu\nu}+\frac{1}{2}m_{\rho}^2\vec{\rho}_{\mu}\cdot\vec{\rho}^{\mu} -\frac{1}{4}F_{\mu\nu}F^{\mu\nu}\ ,  \nonumber \\
\mathcal{L}_{\rm I} & = & \bar{\Psi}\left(-g_{\sigma}\sigma-g_{\omega}\slashed{\omega}-g_{\rho}\vec{\slashed{\rho}}\right)\Psi\ , 
\ee
with $\Psi=\left(\begin{array}{l}p \\ n\end{array}\right)$ and $\psi$ being, respectively, nucleon iso-doublet and electron fields. $A_\mu$ is the EM field, $\sigma, \omega_{\mu}$ and $\vec{\rho}_{\mu}=\rho_{\mu}^i\tau^i$ (with $\tau_i$ being the Pauli matrix) are isoscalar-scalar, isoscalar-vector and isovector-vector meson fields, respectively. The pseudoscalar mesons $\pi$ are neglected since they vanish in the RMF approximation. $\vec{\bm P}_{\mu\nu} = \text D_{\mu}\vec{\rho}_{\nu}-\text D_{\nu}\vec{\rho}_{\mu} - 2ig_\rho\vec{\rho}_{\mu}\times\vec{\rho}_{\nu}$ is the field-strength tensor of rho meson fields. The covariant derivatives are defined as
\be
\text D_{\mu}\Psi & = & \left(\partial_{\mu}+iA_{\mu}Q\right)\Psi
\ , \nonumber\\
\text D_{\mu}\psi & = & \left(\partial_{\mu}-ieA_{\mu}\right)\psi\ ,\nonumber\\
\text D_{\mu}\vec{\rho}_{\nu} & = & \partial_{\mu}\vec{\rho}_{\nu}+iA_{\mu}\left[Q,\vec{\rho}_{\nu}\right]\ ,
\ee
where the charge matrix $Q = e \left(1+\tau_3\right)/2$ with $\tau_3$ being the third component of Pauli matrices.

\begin{figure}[tbh]
    \centering
    \includegraphics[width=1.0\linewidth]{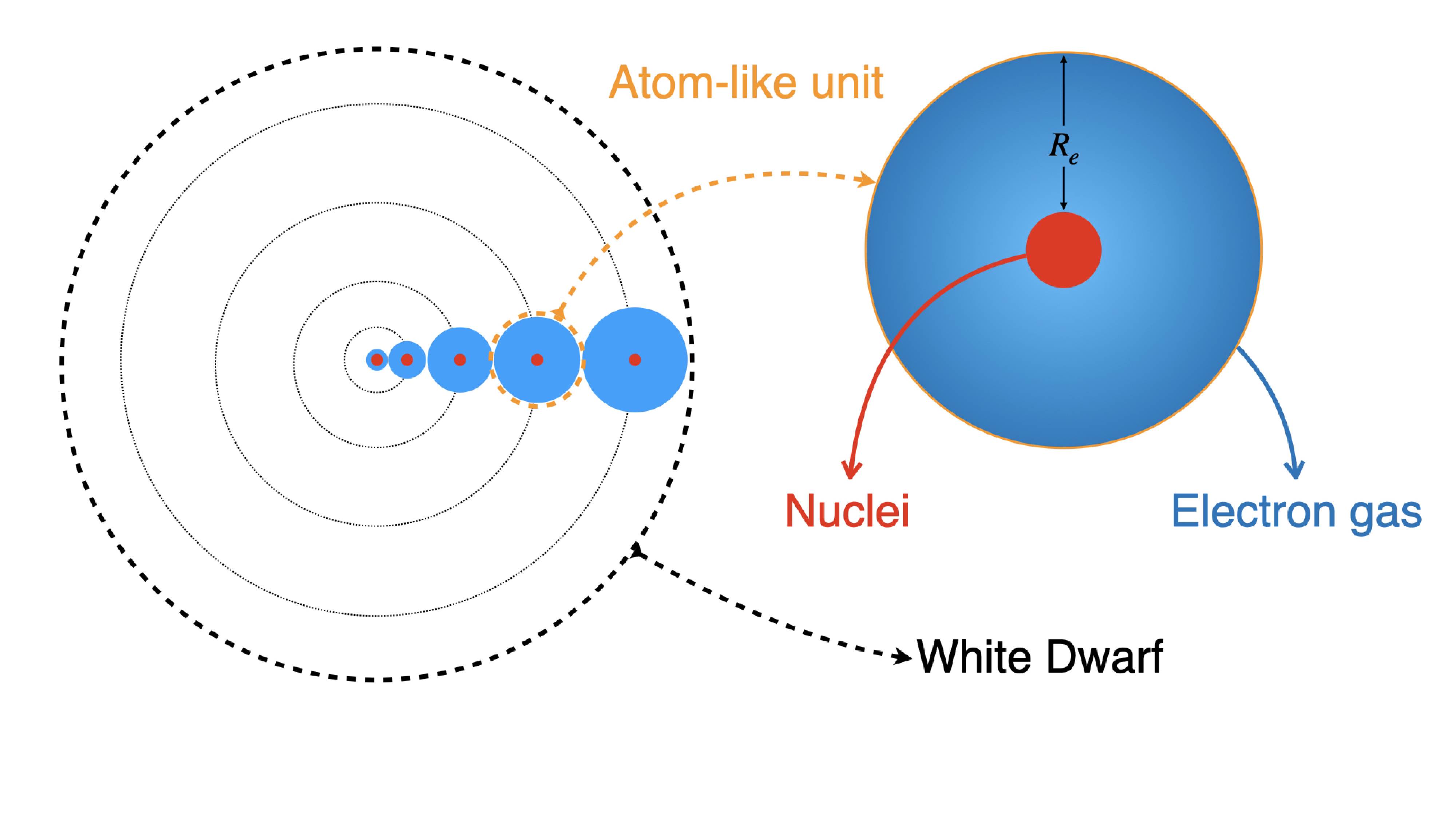}
    \caption{
        The carton of WDs.
        WDs are composed of atom-like units with a single nucleus inside and the nucleus is surrounded by electron gas.
    }
    \label{fig:wd-struc}
\end{figure}

By taking appropriate boundary conditions (BCs), the homogeneous NM and NM clusters can be calculated using the standard RMF approach~\cite{Horowitz:2005zb,Xia:2022dvw}.  However, these BCs cannot be naively extended to the WD matter, as the repulsive force from the EM interaction is much stronger than the attractive nuclear force. Here, we propose that the WD is universally made of atom-like units---Wigner Seitz cells---each containing a nucleus at its core, surrounded by homogeneous electron gas, as illustrated in Fig.~\ref{fig:wd-struc}. The nucleus is calculated by using the microscopic Walecka-type model~\eqref{eq:lag} therefore has a realistic internal structure. The electromagnetic force is obtained by solving the Maxwell equation with a proton source. For simplicity, we assume that both WD and its constituent units are spherically symmetric.

Under the above conditions, the homogeneous condition simplifies to isotropy, allowing us to neglect the time derivatives and spatial components of the meson fields. Furthermore, after applying the RMF approximation, only the neutral $\rho$ meson survives. Consequently, the equations of motion (EOMs) that need to be solved are:
\be
\label{eq:meson-eom}
{}-\nabla^2 A & = & e n_p-e n_e\ ,\nonumber\\
\left({}-\nabla^2+m_\rho^2\right) \rho & = & g_\rho\left(n_p-n_n\right)\ , \nonumber\\
\left({} -\nabla^2+m_\omega^2\right) \omega & = & g_\omega\left(n_p+n_n\right)+c_3\omega^3\ , \nonumber\\
\left({}-\nabla^2+m_\sigma^2\right) \sigma & = & -g_\sigma \left(n_n^s+n_p^s\right)-g_2 \sigma^2-g_3 \sigma^3\ ,
\ee
where $n_{i}$ and $n_{i}^{s}$ are, respectively, the number density and scalar density of particle ``$i$", and the scalar density
\be
n_{n(p)}^s & =& \frac{m_N^{* 3}}{\pi^2}\left[\frac{1}{2}\left(t_{n(p)} \sqrt{1+t_{n(p)}^2}-\operatorname{arcsinh} t_{n(p)}\right)\right]\ ,
\ee
with \(t_{n(p)}=\frac{\left(3 \pi^2 n_{n(p)}\right)^{1 / 3}}{m_N^*}\) and \(m_N^*=m_N+g_{\sigma}\sigma\) defined by EOMs of fermions:
\be
\label{eq:fermion-eom}
& & \left(i\slashed{\partial}-m_N-g_{\sigma}\sigma-g_{\rho}\rho\gamma^0\tau^3-g_{\omega}\omega\gamma^0-A\gamma^0Q\right)\Psi=0\ ,\nonumber\\
& & \left(i\slashed{\partial}-m_e+eA\right)\psi=0\ .
\ee
Note that in Eq.~\eqref{eq:meson-eom} only the zeroth component of the vector meson fields are considered, and the component indices are omitted. By using the EOMs~\eqref{eq:meson-eom} and \eqref{eq:fermion-eom}, the Hamiltonian (energy) density can be obtained via a straightforward Legendre transformation from~\eqref{eq:lag}. It can be decomposed as
\be
\label{eq:energy}
\mathcal{E} & = & \mathcal{E}_{\rm N}+\mathcal{E}_{e}+\mathcal{E}_{\rm NM}+\mathcal{E}_{\rm p \gamma}+\mathcal{E}_{\rm e \gamma}+\mathcal{E}_{\gamma}+\mathcal{E}_{M},
\ee
where $\mathcal{E}_{\rm N},\ \mathcal{E}_{\rm e},\ \mathcal{E}_{\rm NM},\ \mathcal{E}_{\rm p\gamma},\ \mathcal{E}_{\rm e\gamma},\ \mathcal{E}_{\gamma}$ and $\mathcal{E}_{\rm M}$ represent, respectively, the contributions from free nucleon with medium modified mass $m_N^\ast$, free electron, nucleon-meson interaction, proton-photon interaction, electron-photon interaction, pure photon kinetic energy and pure meson interaction. Explicitly, 
\be
\mathcal{E}_{\rm N} & = & \sum_{i=n,p}\frac{1}{8}\left[t_i\sqrt{1+t_i^2}\left(1+2t_i^2\right)-\operatorname{arcsinh} t_i\right] ,\nonumber\\
\mathcal{E}_{\rm e} & = & \frac{1}{8}\left[t_e\sqrt{1+t_e^2}\left(1+2t_e^2\right)-\operatorname{arcsinh} t_e\right]\ ,\nonumber\\
\mathcal{E}_{\rm NM} & = & \Psi^{\dagger}\left(g_{\omega}\omega+g_{\rho}\rho\right)\Psi\ , \nonumber\\
\mathcal{E}_{\rm p\gamma} & = & A\Psi^{\dagger} Q\Psi\ , \nonumber\\
\mathcal{E}_{\rm e\gamma} & = &-\psi^{\dagger}eA\psi\ ,\nonumber\\
\mathcal{E}_{\gamma} & = &{} -\frac{1}{2}\left(\nabla A\right)^2\ ,\nonumber\\
\mathcal{E}_{\rm M} & = &{} - \frac{1}{2}\left(\nabla\sigma\right)^2-\frac{1}{2}m_{\sigma}^2\sigma^2+\frac{1}{3}g_2\sigma^3+\frac{1}{4}g_3\sigma^4 \nonumber\\
&  &{} - \frac{1}{2}\left(\nabla\omega\right)^2-\frac{1}{2}m_{\omega}^2\omega^2+\frac{1}{4}c_3\omega^4\nonumber\\
&  &{} -\frac{1}{2}\left(\nabla\rho\right)^2-\frac{1}{2}m_{\rho}^2\rho^2\ .
\ee

To study the structure of nuclei within the atom-like units using model~\eqref{eq:lag}, we adopt the parameter values listed in Tab.~\ref{tab:para} which were obtained by meticulously calibrating to align with the properties of NM and the M-R relations of NSs~\cite{Guo:2023mhf}. The nuclei are assumed to be spherical, and the energy density can be calculated by neglecting the contributions from electrons within each atom-like unit, as a result of Gauss theorem, while still including the the EM interactions between protons. Specifically, contributions $\mathcal{E}_{\rm N}, \mathcal{E}_{\rm NM}, \mathcal{E}_{\rm p \gamma}, \mathcal{E}_{\gamma}$ and $\mathcal{E}_{M}$ from Eq.~\eqref{eq:energy} are considered.
In our calculations, the initial distribution of hadron fields is set within a sphere of radius \(R_0=5\ \rm fm\), which is a typical size of a nucleus. The EOMs~\eqref{eq:meson-eom} are iteratively solved to determine the ground state of corresponding nuclei, which gives us the reasonable geometry information of the corresponding nuclei. The boundary conditions are defined as the nuclear density $n_{n(p)}(r=R_0)=0$ and $n^\prime_{n(p)}(r=R_0)=0$. Explicit check using $R_0=(4-10)$ fm shows that the results of nucleus properties are intact. Our results of the nuclei properties are shown in Tab.~\ref{tab:nuclei}.

\begin{table}[htbp]
	\centering
	\caption {
        Optimal values of the parameters obtained from pinning down NM properties and NS star structures~\cite{Guo:2023mhf}.
    }
	\label{tab:para}
	\begin{tabular}{ccccccc}
		\hline
		\hline
		$g_{\sigma}$ & $g_{\omega}$  & $g_{\rho}$ & $g_3$ & $c_3$ & $g_2$ & $ m_{\sigma}$   \cr
		\hline
		$9.82$ &~ $11.8 $ &~ $3.42$ & ~$1.26$ & ~$72.6$ & ~$-1550$~MeV &~ $531$~MeV \cr
		\hline
		\hline
	\end{tabular}
\end{table}

\begin{widetext}
\begin{center}
\begin{table}[htbp]
	\centering
	\small
	\caption{
		The results of nuclei properties.
        The root-mean-square radius $\sqrt{\langle r^2\rangle}$ is calculated via \(\langle r^2\rangle=\int_0^{R_0} R^2 \rho(R) 4 \pi R^2 {\rm d} R/\int_0^{R_0} \rho(R) 4 \pi R^2 {\rm d} R\), where \(\rho(R)\) is the distribution of nucleons of the ground state.
        The binding energy $\rm B.E.$ is defined by $m_N-E_{i}/(N+Z)$, where $E_i$ is the total energy of an atomic unit and ``$i$" refers to different elements with \(N(Z)\) being the number of neutron (proton) inside the nuclei.
        The experiment values (denoted as ``Exp.") are taken from Refs.~\cite{Angeli:2013epw, Fortune:2016jzj, NNDCNuDat}.
        The radius \(\sqrt{\langle r^2\rangle}\) and B.E. are in units of fm and MeV, respectively.
    }
	\label{tab:nuclei}
	\begin{tabular}{cccccccccccc}
		\hline
		\hline
		\hline
		Elements \;\; & \(^{4}\rm He\)  \;& \(^{12}\rm C\) \;& \(^{13}\rm C\) \;& \(^{14}\rm C\) \;& \(^{15}\rm C\) \;& \(^{16}\rm C\) \;& \(^{17}\rm C\) \;& \(^{18}\rm C\) \;& \(^{19}\rm C\) \;& \(^{20}\rm C\) \;& \(^{16}\rm O\)  \cr
		\hline
        \hline
		$\sqrt{\langle r^2\rangle}$(Exp.)\;\; & - & - & - & $2.33(7)$ & $2.54(4)$ & $2.74(3)$ & $2.76(3)$ & $2.86(4)$ & $3.16(7)$& $2.98(5)$& -  \cr
 		\hline
        $\sqrt{\langle r^2\rangle}$(The.)\;\; & $1.90$ & $2.34$ & $2.36$ & $2.43$ & $2.48$ & $2.55$ & $2.61$ & $2.66$ & $2.71$& $2.77$& $2.51$  \cr
        \hline
        \hline
		B.E. (Exp.)\;\; & $7.07$ & $7.68$ & $7.47$ & $7.52$ & $7.10$ & $ 6.92$ & $ 6.56$ & $6.43$ & $6.12$& $5.96$& $7.98$  \cr
		\hline
        B.E. (The.)\;\; & $3.45$ & $6.43$ & $6.63$ & $6.65$ & $6.61$ & $6.50$ & $6.37$ & $6.20$ & $6.01$& $5.81$& $7.08$ \cr
		\hline
		\hline
		\hline
	\end{tabular}
\end{table}
\end{center}
\end{widetext}

From Tab.~\ref{tab:nuclei}, one can see that our current approach successfully captures the qualitative properties of nuclei, with the exception of the binding energy of \(^{4}\rm He\), which is unnaturally deep bounded~\cite{NPLQCD:2012mex}.
Furthermore, the results increasingly diverge from experiment data with \(N-Z\).
This deviation is attributed to the oversimplified parametrization of isospin-related interactions, especially those involving the \(\rho\) meson.

\section{Structure of white dwarfs and binary inspirals}
\label{sec:wd}

The energy of an atom-like unit consists of the contributions from the nucleus core, which is the nucleus mass calculated above, as well as the energy of free electrons, free photons, and the Coulomb energy between the electrons and the nucleus core. By varying the volume of the unit, we can obtain both the the number density \(\mathcal{N}\) and the energy density $\mathcal{E}$ of the atom-like units. The pressure of the unit can be derived through the thermal relation
\begin{equation}
    P={}-\mathcal{E}+\mathcal{N} \frac{\mathrm{d} \mathcal{E}}{\mathrm{d} \mathcal{N}}\ ,
\end{equation}
It should be noted that the pure mesonic contributions to the energy density vanish outside the nuclei, while the electron density disappears within the nucleus. Our numerical results of the nucleon density distributions of different nuclei and the corresponding EOS of WD matter are presented in App.~\ref{sec:appEOS}.

\subsection{Structure of white dwarf}

After obtaining \(\mathcal{E}-P\) relation, the M-R relations can be calculated by solving the Tolman-Oppenheimer-Volkoff (TOV) equations~\cite{Tolman:1939jz,Oppenheimer:1939ne} and the TD of WDs can also be obtained by solving the Love number equation following Ref.~\cite{Flanagan:2007ix, Postnikov:2010yn}. It should be noted that in our current work, we will not consider the effect of $\beta$-equilibrium~\cite{Salpeter:1961zz,Rotondo:2009cr,Rotondo:2011zz,Boshkayev:2012bq}.

\begin{figure}[htbp]
    \centering
    \includegraphics[width=1.0\linewidth]{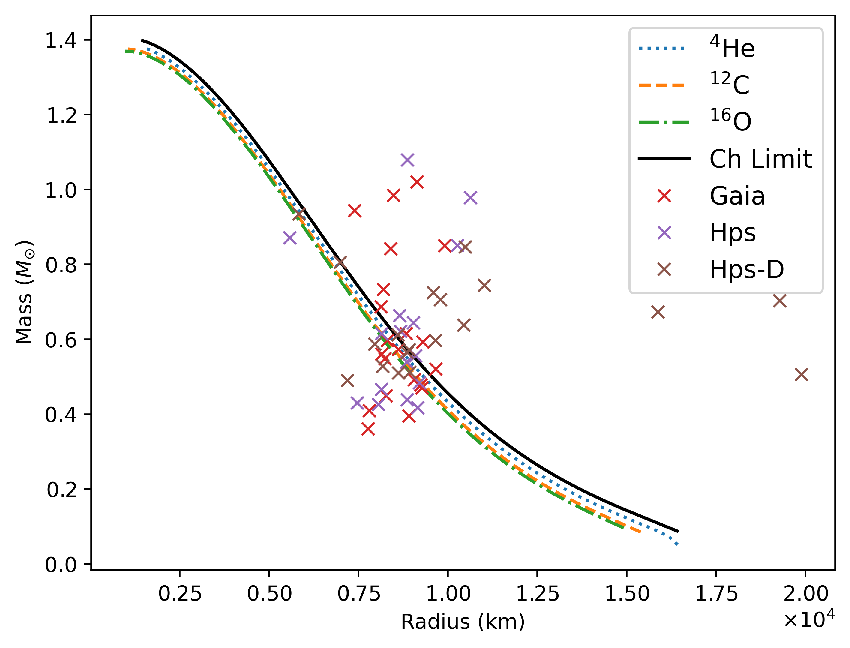}
    \includegraphics[width=1.0\linewidth]{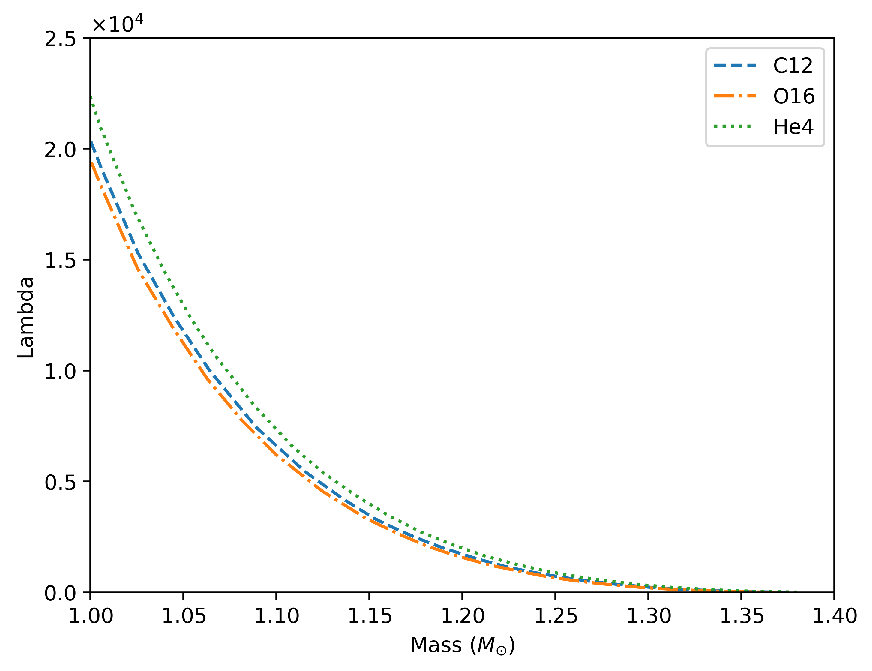}
    \caption{
        {The M-R relation (upper panel) and TD (lower panel) of WDs. The observational data marked by "$\times$" are taken from Ref.~\cite{tremblay2016gaia}. The Chandrasekhar limit of our parametrization (denoted as ’Ch’) is also plotted for comparison.}
    }
    \label{fig:WD-struc}
\end{figure}

Our results are resented in Fig.~\ref{fig:WD-struc}, where it can be seen that our WD structures saturate the regular Chandrasekhar limit (free electron gas, a pivotal concept in astrophysics)~\cite{Chandrasekhar:1931ih}.
A comparison of the pressure at different mass density obtained from our full calculation (\(P_{\rm RMF}\)) and free electron limit (\(P_{\rm RMF}^{\rm Ch}\)), Chandrasekhar limit ($P_{\rm Ch}$)~\cite{Chandrasekhar:1931ih}, Salpeter approach ($P_{\rm S}$)~\cite{Salpeter:1961zz}, and the relativistic FMT from Ref.~\cite{Rotondo:2011zz} (\(P_{\rm FMT}^{\rm rel}\)) is shown in Table~\ref{tab:chlimit}.
It can be seen that, in our approach, the EOS of WDs deviates slightly more from the Chandrasekhar limit compared to the Salpeter approach and the relativistic FMT framework, both of which provide corrections to the WD structures at fine structure level.

\begin{widetext}
\begin{center}
\begin{table}[htbp]
	\centering
	\caption {
		{
			The pressure at different mass densities \(\rho_{\rm mass}\) for regular Chandrasekhar limit (\(P_{\rm Ch}\))~\cite{Chandrasekhar:1931ih}, Salpeter approach (\(P_{\rm S}\))~\cite{Salpeter:1961zz}, the relativistic FMT (\(P_{\rm FMT}^{\rm rel}\))~\cite{Rotondo:2011zz}, our parametrization with free electron gas limit (\(P_{\rm RMF}^{\rm Ch}\)), and our parametrization (\(P_{\rm RMF}\)).
			\(\rho_{\rm mass}=E\cdot \rho_{\rm atoms}\) is unit of \(\rm g/cm^3\), where $E$ is the total energy of every atomic unit and $\rho_{\rm atoms}$ is the number density of atom-like units, and pressure is in the unit of \(\rm dyne/cm^2\). We take $^{12}$C as an example.
		}
	}
	\label{tab:chlimit}
	\begin{tabular}{cccccc}
		\hline
		\hline
		$\rho_{\rm mass}$ & $P_{\rm Ch}$~\cite{Chandrasekhar:1931ih}  & $P_{\rm S}$~\cite{Salpeter:1961zz} & $P_{\rm FMT}^{\rm rel}$~\cite{Rotondo:2011zz} & $P_{\rm RMF}^{\rm Ch}$ & $ P_{\rm RMF}$   \cr
		\hline
		$10^{4}$ &~ $1.45\times10^{19} $ &~ $1.29\times10^{19}$ & ~$1.29\times10^{19}$ & ~$1.45\times10^{19}$ & ~$1.27\times10^{19}$  \cr
		\hline
		$10^{5}$ &~ $6.50\times10^{20} $ &~ $6.14\times10^{20}$ & ~$6.13\times10^{20}$ & ~$6.48\times10^{20}$ & ~$6.06\times10^{20}$  \cr
		\hline
		$10^{6}$ &~ $2.63\times10^{22} $ &~ $2.55\times10^{22}$ & ~$2.54\times10^{22}$ & ~$2.62\times10^{22}$ & ~$2.52\times10^{22}$  \cr
		\hline
		$10^{7}$ &~ $8.46\times10^{23} $ &~ $8.29\times10^{23}$ & ~$8.27\times10^{23}$ & ~$8.44\times10^{23}$ & ~$8.20\times10^{23}$  \cr
		\hline
		$10^{8}$ &~ $2.15\times10^{25} $ &~ $2.11\times10^{25}$ & ~$2.11\times10^{25}$ & ~$2.14\times10^{25}$ & ~$2.09\times10^{25}$  \cr
		\hline
		$10^{9}$ &~ $4.86\times10^{26} $ &~ $4.78\times10^{26}$ & ~$4.77\times10^{26}$ & ~$4.84\times10^{26}$ & ~$4.74\times10^{26}$  \cr
		\hline
		$10^{10}$ &~ $1.06\times10^{28} $ &~ $1.04\times10^{28}$ & ~$1.04\times10^{28}$ & ~$1.05\times10^{28}$ & ~$1.03\times10^{28}$  \cr
		\hline
		\hline
	\end{tabular}
\end{table}
\end{center}
\end{widetext}

\begin{figure}[htbp]
	\centering
	\includegraphics[width=1.0\linewidth]{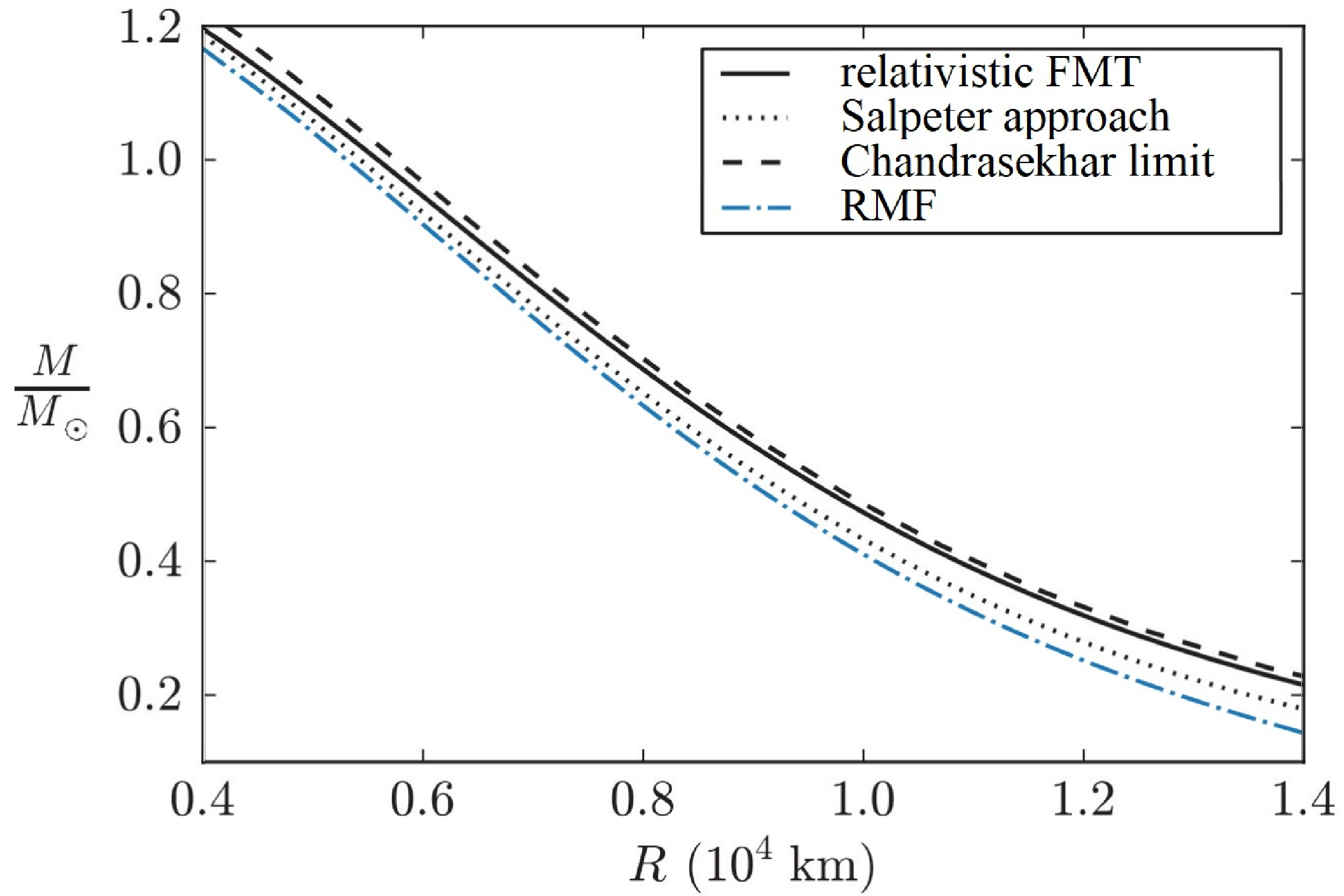}
	\caption{
			Comparison of the M-R relations of \(^{12}\rm C\) WD calculated from our model (RMF), relativistic FMT~\cite{Rotondo:2011zz}, Salpeter approach~\cite{Salpeter:1961zz} and Chandrasekhar limit~\cite{Chandrasekhar:1931ih}.
			It should be noted that the M-R relation of relativistic FMT is calculated by using higher order TOV equations, while others are calculated by using the standard TOV equations.
	}
	\label{fig:compareMRC}
\end{figure}

We compare in Fig.~\ref{fig:compareMRC} the M-R relations calculated from our model (RMF) and relativistic FMT~\cite{Rotondo:2011zz}, Salpeter approach~\cite{Salpeter:1961zz} and Chandrasekhar limit~\cite{Chandrasekhar:1931ih}.
Our results align well within the region of the observational data, thereby validating the effectiveness of our theoretical framework in describing the properties of WDs. Furthermore, a realistic description of the nuclei within WDs will yield corrections of a similar magnitude to those arising from considerations of electromagnetic interactions or relativistic fluid corrections in WD matter. This signifies the importance of including the nucleus structure in the WD study.

\subsection{Gravitational waves of binary inspirals}

\begin{figure}[htbp]
    \centering
    \includegraphics[width=1.0\linewidth]{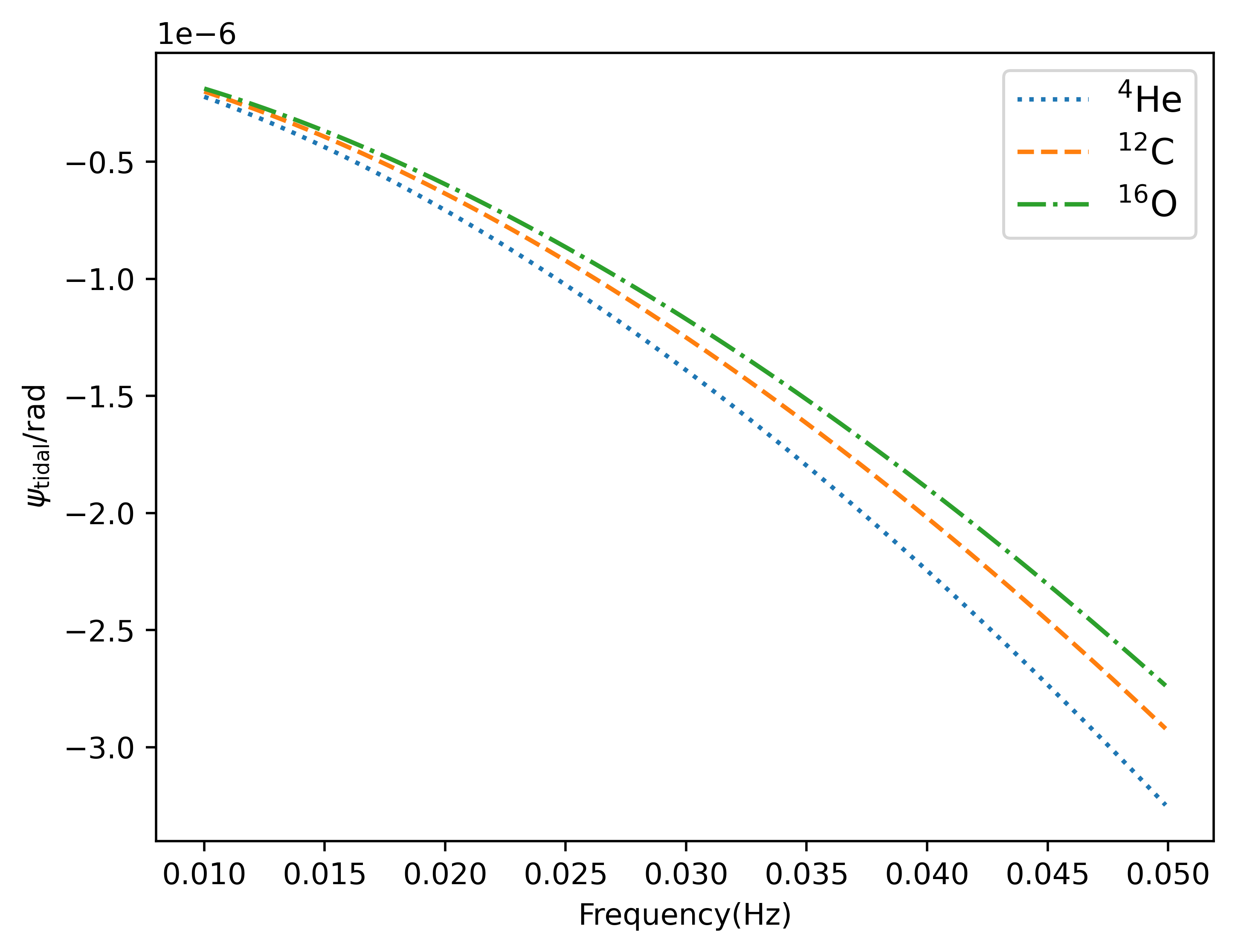}
    \includegraphics[width=1.0\linewidth]{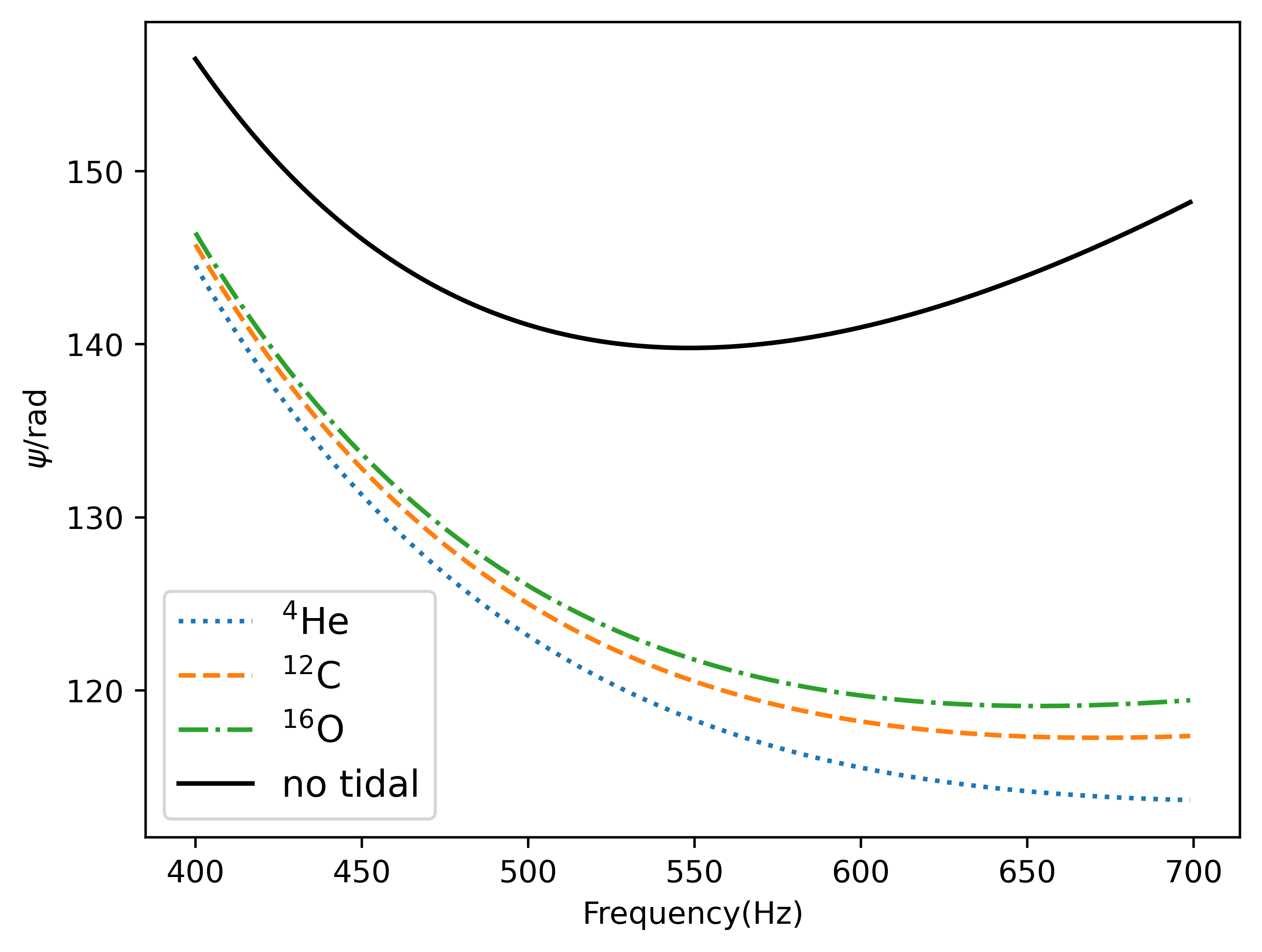}
    \caption{
        The phase shift of GWs in mHz region (upper panel) and hundred Hz region (lower panel). 
    }
    \label{fig:phase-f}
  \end{figure}

\begin{figure}[htbp]
    \centering
    \includegraphics[width=1.0\linewidth]{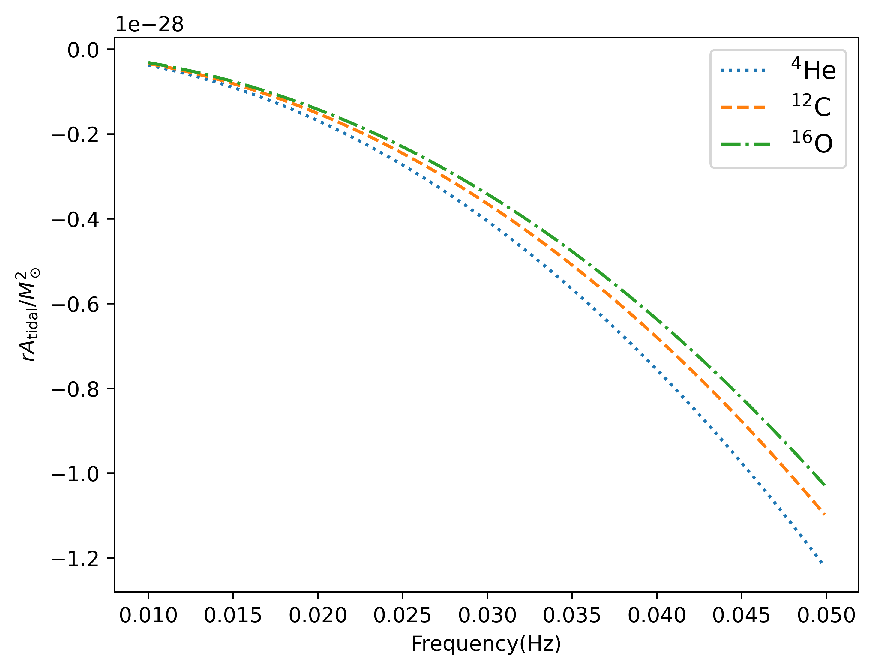}
    \includegraphics[width=1.0\linewidth]{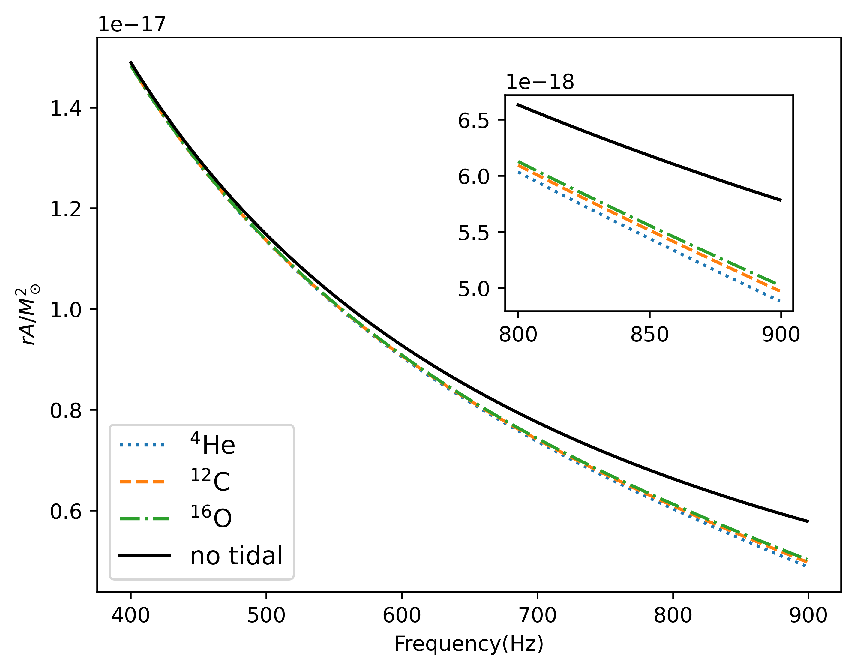}
    \caption{
        The amplitude deviation of GWs in mHz region (upper panel) and hundred Hz region (lower panel) introduced by TD. 
    }
    \label{fig:amp-f}
\end{figure}

We finally calculate the GW signals from the binary WD inspirals. In Ref.~\cite{ZeCheng:2020dtf}, the GW signals emitted from WD coalescences were studied in the Newtonian limit. Here, we will investigate the signals with and without tidal deformability at the 2.5PN order. The expression of the GW waveforms up to the 2.5PN order are given in App.~\ref{sec:appPN}. The results of the GWs from binary WDs composed of different elements in frequency-domain, are shown in Fig.~\ref{fig:phase-f} and Fig.~\ref{fig:amp-f} for, respectively, the phase shift and amplitude deviation. We typically choose the equal mass system with component mass $1.1\ M_{\odot}$. The corresponding tidal deformability are \(\Lambda=7352,\ 6613,\ 6197\) for $^{4}\rm He$, $^{12}\rm C$, and $^{16}\rm O$, respectively, according to the results shown in Fig.~\ref{fig:WD-struc}.

Our results, as illustrated in Fig.~\ref{fig:phase-f} and Fig.~\ref{fig:amp-f}, indicate that WDs composed of lighter elements exhibit a bigger deviation from the point-particle approximation---the GW signals without TD---in terms of both phase and amplitude, in line with the tidal deformability shown in Fig.~\ref{fig:WD-struc}. Although the PN approximation may not be suitable for the merging stage of WD inspirals, particularly in the hundred Hz frequency range, the discernible differences between the point particle approximation and the tidal deformability corrections offer a preliminary understanding of matter effects on GWs. This insight is especially valuable given the complexities involved in numerical simulations for such systems. 
Comparative analyses in both the mHz and hundred Hz regions could serve as initial guides for future GW detections and as a testbed for further research on WDs.
Additionally, understanding the limitation of point particle approximation in GW physics is crucial for interpreting GW signals.

\begin{figure}[htbp]
	\centering
	\includegraphics[width=1.0\linewidth]{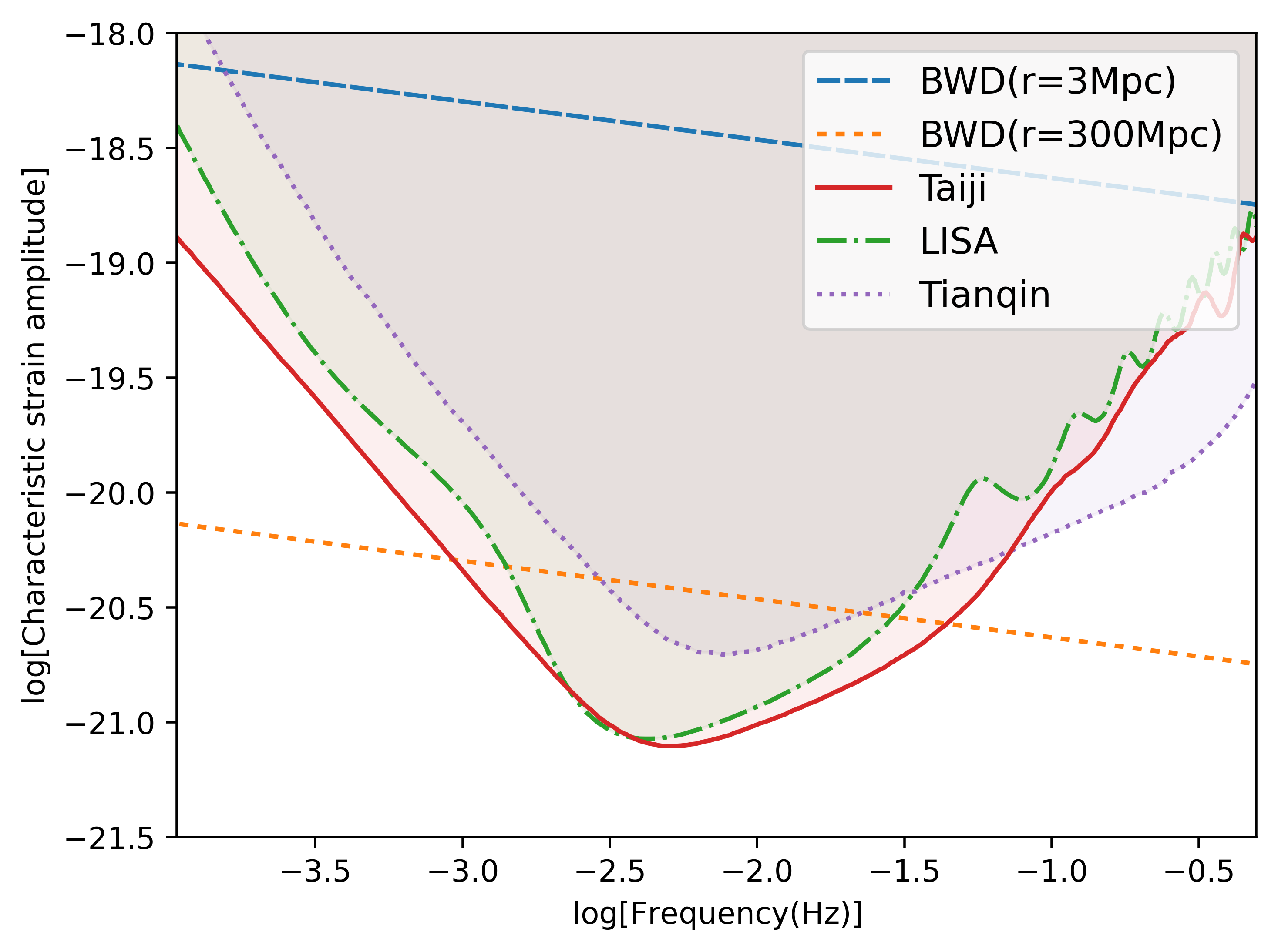}
	\caption{
		The predicted characteristic strain amplitude of GWs in comparison with the detectabilities of space-based facilities.
	}
	\label{fig:GWBound}
\end{figure}

In addition, it is interesting to study the dimensionless characteristic strain amplitude $h_c(f)$~\cite{Moore:2014lga}
\be
h_c(f) & = & \sqrt {f S_h(f)}=2f|\tilde h(f)|
\ee
where $S_h(f)$ is the power spectral density (PSD). Our results in comparison with the detectabilities of LISA~\cite{LISACosmologyWorkingGroup:2019mwx}, Taiji~\cite{LUO2020102918,Luo:2021qji}, and TianQin~\cite{Gong:2021gvw}, at $r=3$~Mpc and 300~Mpc, are presented in Fig.~\ref{fig:GWBound}. With the result of characteristic strain $h_c(f)$, we calculate the signal-to-noise ratio (SNR) $\varrho$ defined as~\cite{Moore:2014lga, Yin:2023gwc}
\begin{equation}
	\varrho=\sqrt{\int^{f_{\max}}_{f_{\min}} \frac{4|\tilde h(f)|^2}{S_n(f)} \mathrm df}
	=\sqrt{\int^{\log f_{\max}}_{\log f_{\min}} \left(\frac{h_c(f)}{h_n(f)}\right)^2 \mathrm d(\log f)},
\end{equation}
where $S_n$ is the one-sided noise PSD, and $h_n(f)=\sqrt{fS_n(f)}$. In the frequency range 0.0420-0.0428 Hz within the Taiji's mission duration $\sim5$ years, $\varrho\sim10$ for $r=3$~Mpc and $\varrho\sim0.1$ for $r=300$~Mpc. One can see that, as expected, the events of the inspirals at $\lesssim 3$~Mpc are indeed detectable by these space-based facilities.

\section{Summary and perspective}
\label{sec:out}

In this work, we successfully extend the RMF approach to study the WDs by using a Walecka-type model that incorporates hadrons, electrons and photons in a unified framework. We propose that a WD is composed universally of a single element atom-like constituents---Wigner-Seitz cells. The properties of the nuclei and WDs obtained from our model are closely aligned with observational data. Additionally, we investigate the GWs stemming from the merger of binary WDs.

Although estimates of nucleus structure or corresponding physical process in WD studies can be simply obtained through experimental data, we believe that the approach established in this work is particularly valuable for studing WDs composed of elements, whose structures are not well measured. Additionally, our framework is usefully for examining the process of continuous neutronization (beta equilibrium) in WDs and NS crusts with the density changing. That's because an advantage of our approach is its ability to describe NSs (homogeneous nuclear matter), NS crusts (cluster of nuclear matter), and WDs (nucleus surrounded by electron gas) within a single, cohesive framework by employing distinct BCs.

Based on the parameters established from the properties of ordinary homogeneous NM at saturation density, we estimated the structure of nuclei. The results obtained are closely aligned with experiment data~\cite{Fortune:2016jzj,NNDCNuDat}, except for the anomalously deep bounded \(^{4}\rm He\) and nuclei with large neutron-proton (\(N-Z\)) differences. This discrepancies are attributed to the simplified interaction parametrizations and ignorance of the shell effect. 
Subsequently, we computed the M-R relations and TD of WDs, conceptualizing them as an atom-like unit composed objects. We found that the yielded results are in concordance with observational data~\cite{tremblay2016gaia}. Our approach incorporates the size effect of nucleus, Coulomb force and photon energy in a unified model. The obtained results show a more obvious fine structure deviation compared to other methods. This congruence underscores the efficacy of our model in contributing to the understanding of the WD physics at a fine structure level.

Finally, we generated GW signals for binary WD inspirals composed of various elements, including \(^{4}\rm He\), \(^{12}\rm C\), and \(^{16}\rm O\), considering both scenarios with and without TD across frequency ranges from mHz to hundred Hz, up to the 2.5 PN order. The resulting signals indicate that WDs composed of lighter elements exhibit bigger deviations from the point-particle approximation, which is consistent with the finding related to TD. These results provide valuable preliminary guidelines for interpreting future GW detection signals.

We should say that the model used and the structure of WDs considered in this work are immature. Several improvements deserve further consideration:

Firstly, as is well known, the present used Walecka-type model is insufficient for accurately describing nuclei and NM. Therefore, it would be helpful and interesting to refine the calculations using more realistic models based on the chiral symmetry of QCD~\cite{Jenkins:1990jv,Bernard:2007zu,Scherer:2009bt,Ma:2023eoz,Li:2016uzn,Ma:2016gdd}, ab initio parameterized nuclear force~\cite{Bender:2003jk,Stoitsov:2010ha,Hergert:2015awm,Duguet:2015nna} or Skyrme force parametrization~\cite{Agrawal:2006ea,Lesinski:2006cu,Zuo:2017njs}. We expect that these studies will enhance our understanding of the interplay between nuclear structure physics and QCD, allowing us to access nuclear structure with greater precision.

Secondly, we have so far assumed that WDs are composed of the atom-like units with a fixed nucleus. To take the shell structure of WDs into account, the transition between elements should be properly considered in the future work.

Thirdly, although this preliminary work captures the dominant physics of WD, we have only considered WDs composed of a unique nucleus. It is accepted that the inverse beta-decay of nucleus plays a fundamental role in WD physics~\cite{Rotondo:2009cr,Rotondo:2011zz,Boshkayev:2012bq}. This aspect deserves a serious consideration within the current framework.

Last but not least, since binary WDs may get crystallized after a long evolution, it is interesting to consider how the crystallization of WDs affects their tidal deformations. Such changes could also contribute to the GWs generated during the inspiral and merger of binary WDs or binary white dwarf-neutron star systems. These phenomena are detectable by space-based facilities as discussed in~\cite{Tang:2022kmn, Perot:2022zwy}. We leave this topic to our future work.

\section*{ACKNOWLEDGMENTS}
Y.~L. M. would like to thank C. J. Xia for his valuable discussion. The work of Y.~L. M. was supported in part by the National Key R\&D Program of China under Grant No. 2021YFC2202900 and the National Science Foundation of China (NSFC) under Grant No. 12347103 and No. 11875147.
Y.~L.~W. was supported in part by the National Key Research and Development Program of China
under Grant No.2020YFC2201501, the National Science Foundation of China (NSFC) under Grants No. 12347103, No. 12147103, No. 11821505, and the Strategic Priority Research Program of the Chinese Academy of Sciences under Grant No. XDB23030100.

\appendix


\begin{widetext}
	
\section{The nuclei structures and EOS of white dwarfs}
\label{sec:appEOS}

Our results of the nucleon density distribution in an isotope and EOS of WDs are plotted, respectively, in Fig.~\ref{fig:nuclei} and Fig.~\ref{fig:EOS}.
\begin{figure}[htbp]
	\centering
	\subfigure[\(^{4}\rm He\)]{\includegraphics[scale=0.6]{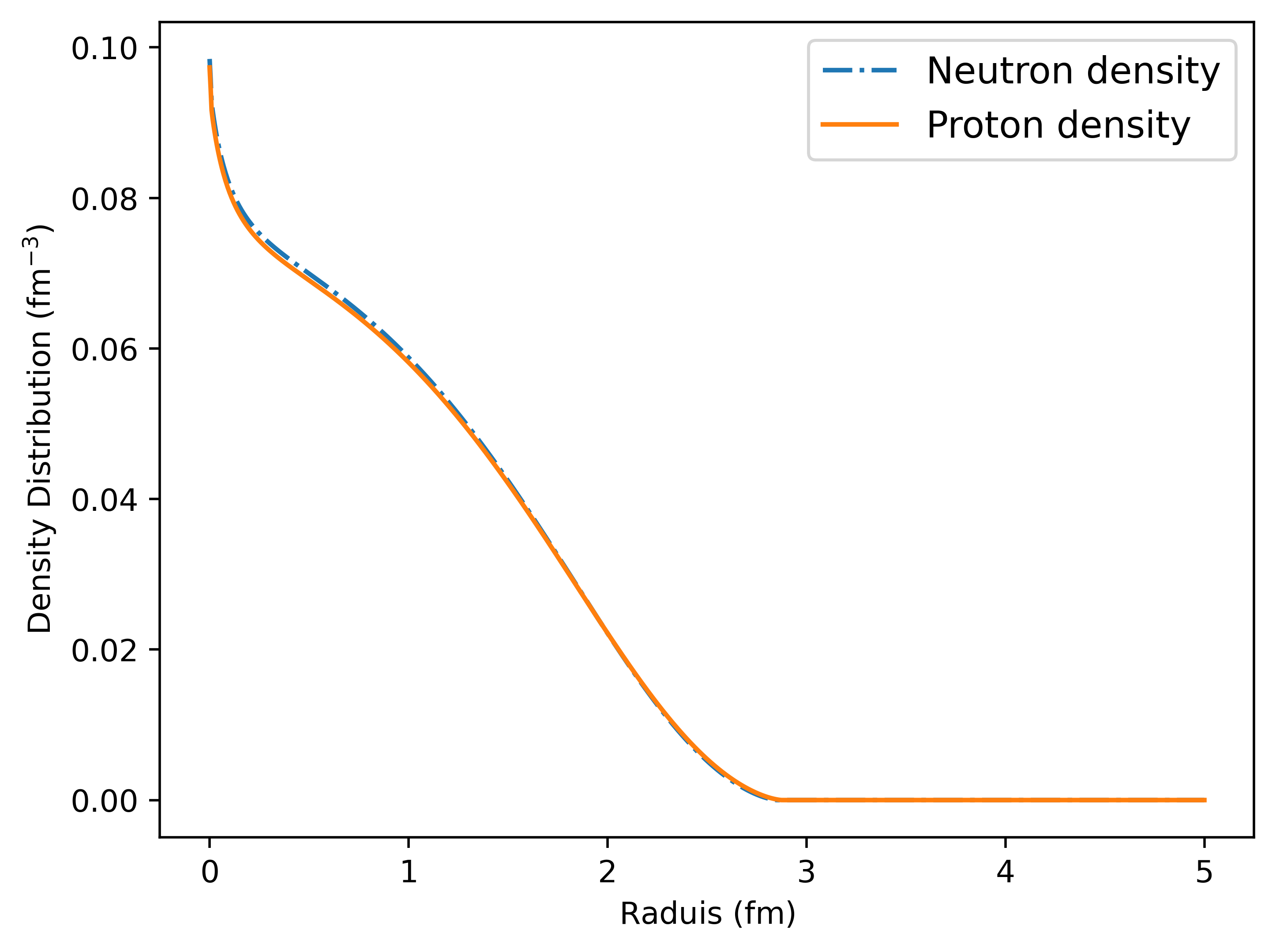}} 
	\subfigure[\(^{16}\rm O\)]{\includegraphics[scale=0.6]{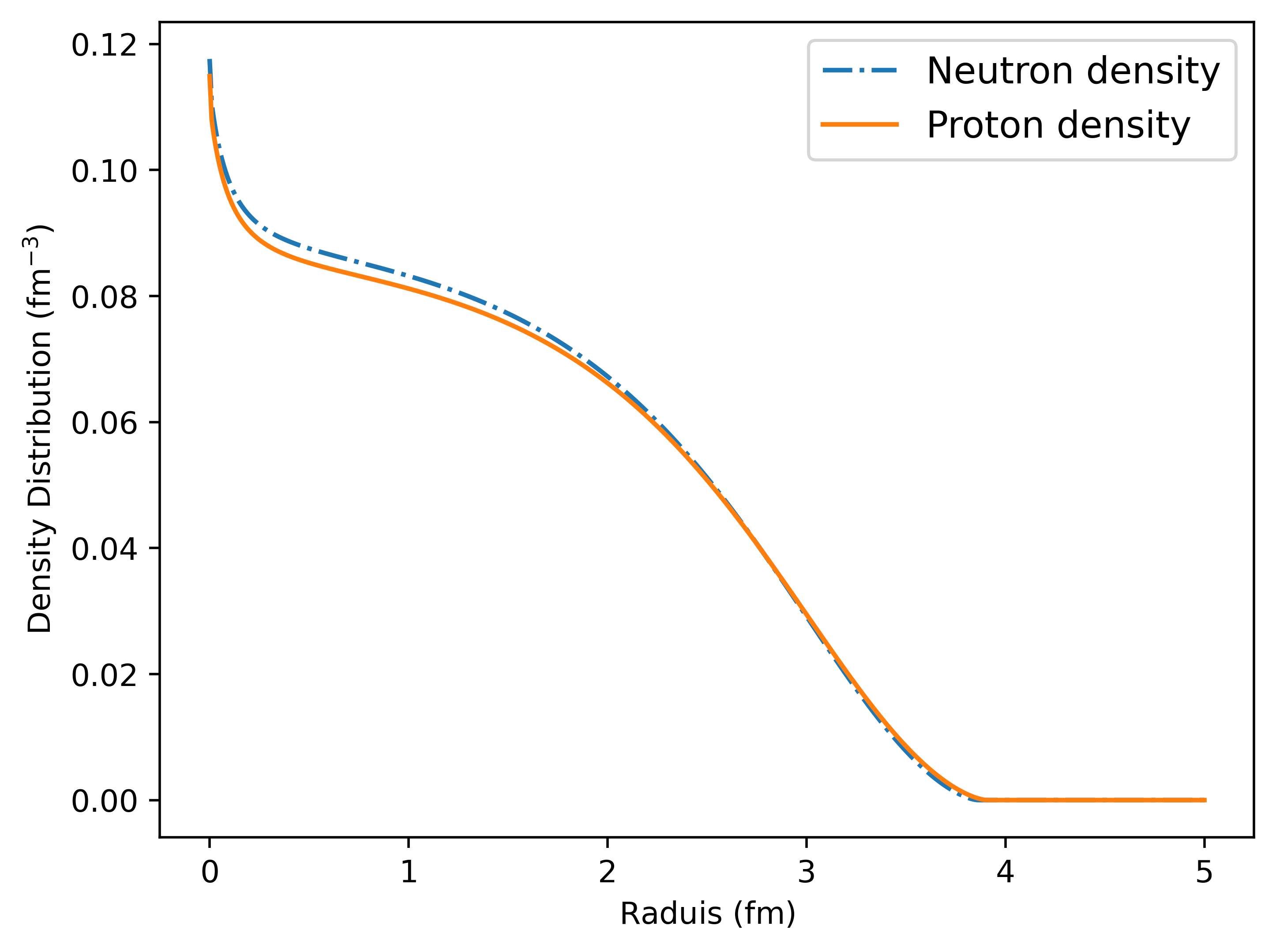}}
	\subfigure[\(^{12}\rm C\)]{\includegraphics[scale=0.6]{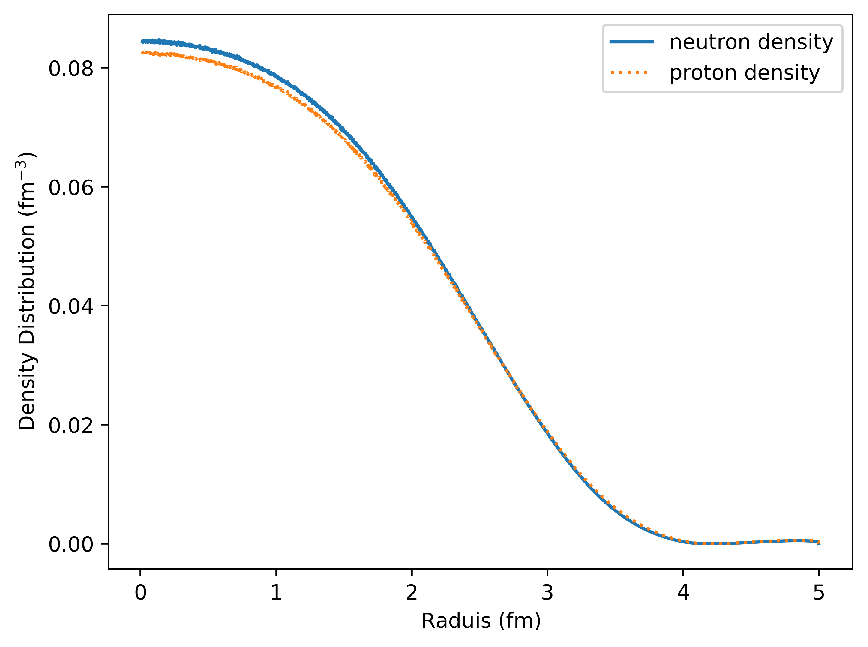}}
	\subfigure[\(^{13}\rm C\)]{\includegraphics[scale=0.6]{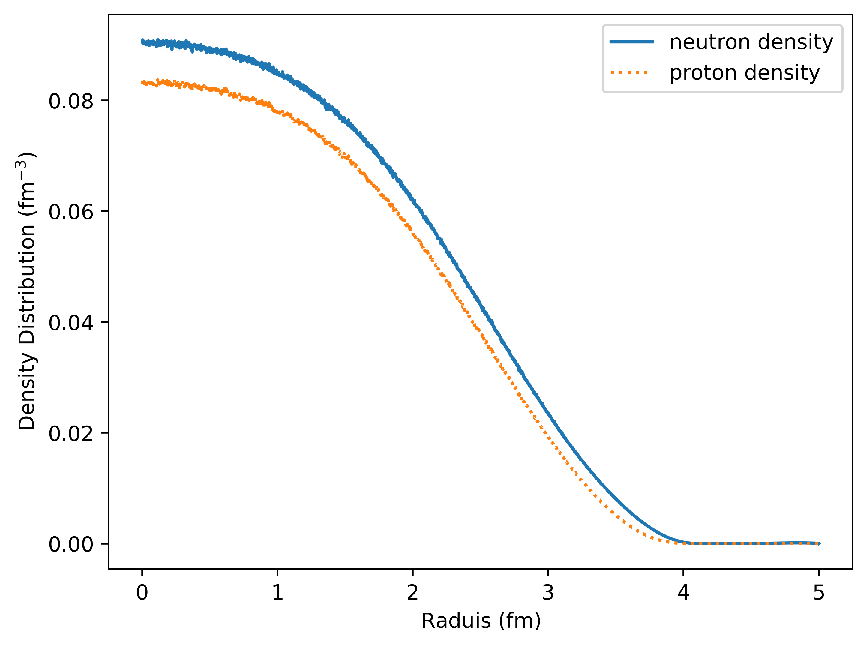}}
	\subfigure[\(^{14}\rm C\)]{\includegraphics[scale=0.6]{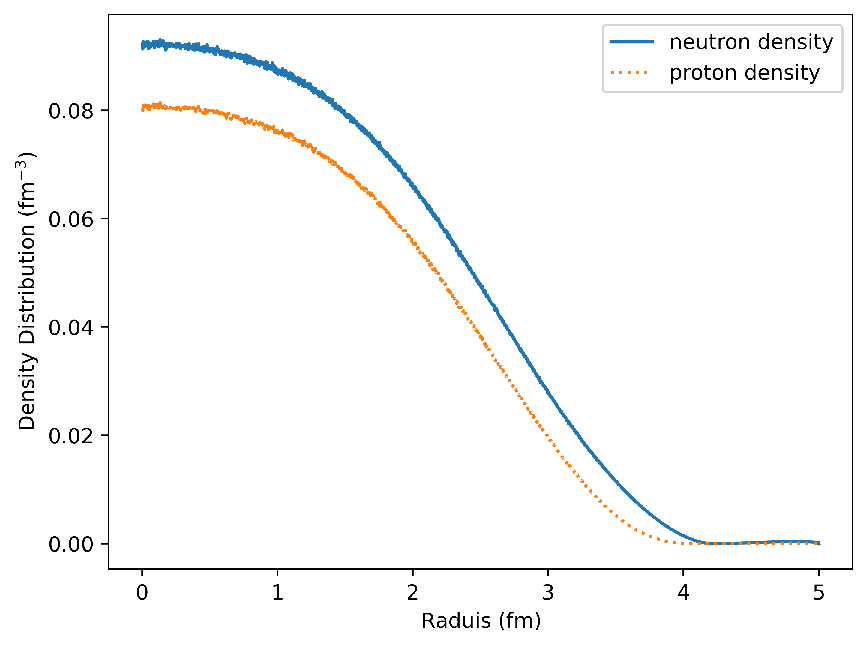}}
	\subfigure[\(^{15}\rm C\)]{\includegraphics[scale=0.6]{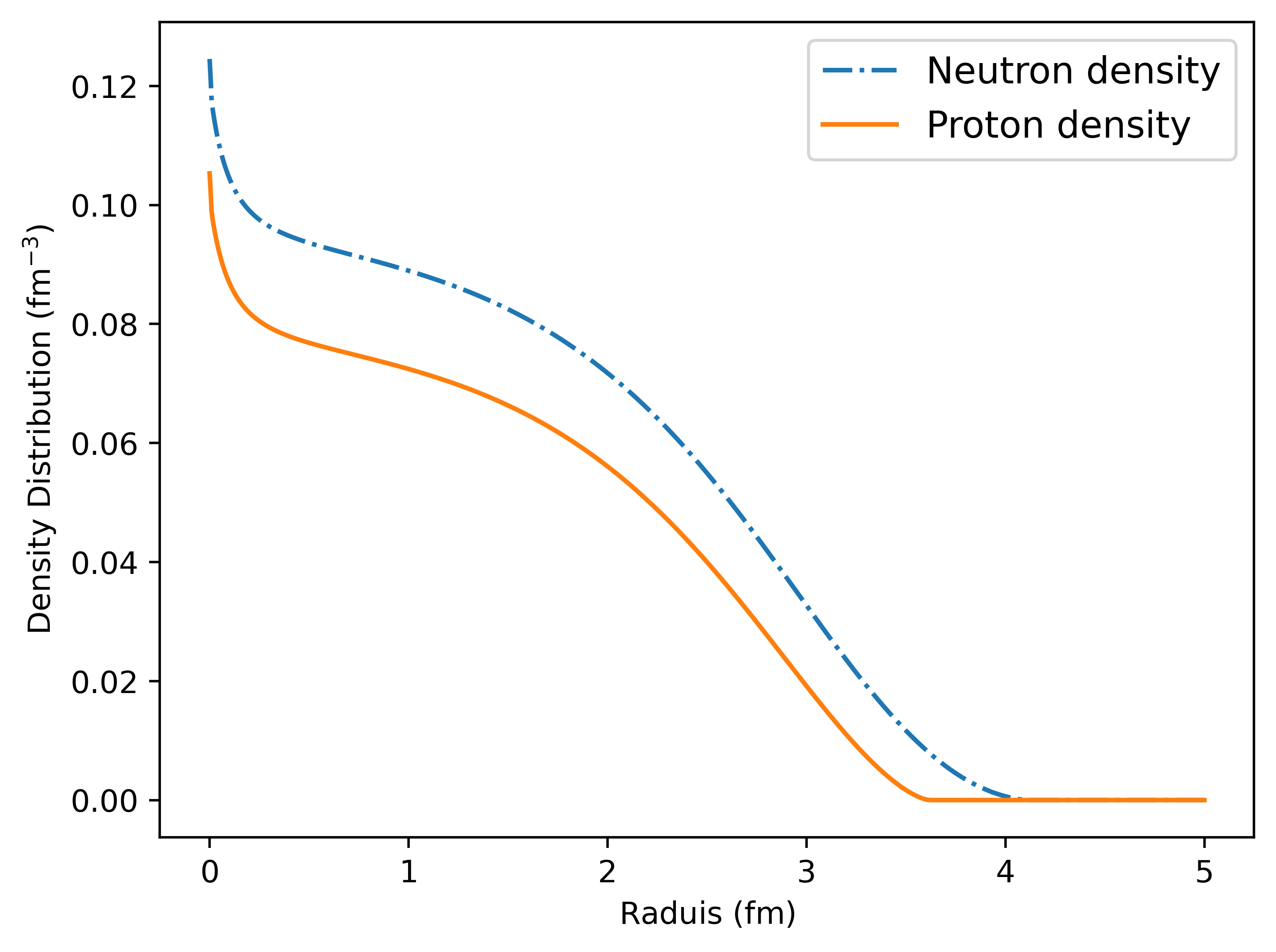}}
\end{figure}
\begin{figure}[htbp]
	\ContinuedFloat
	\centering
	\subfigure[\(^{16}\rm C\)]{\includegraphics[scale=0.6]{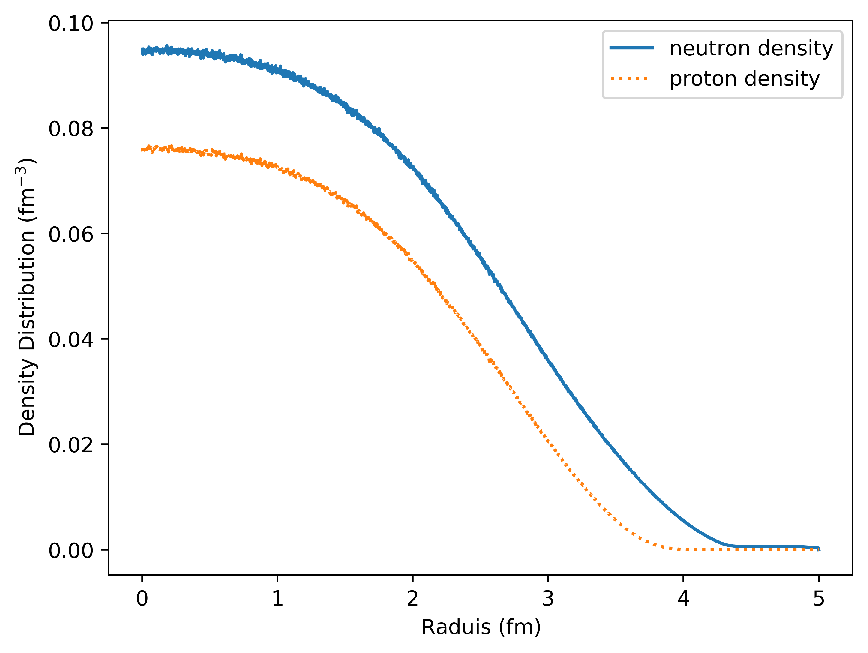}}
	\subfigure[\(^{17}\rm C\)]{\includegraphics[scale=0.6]{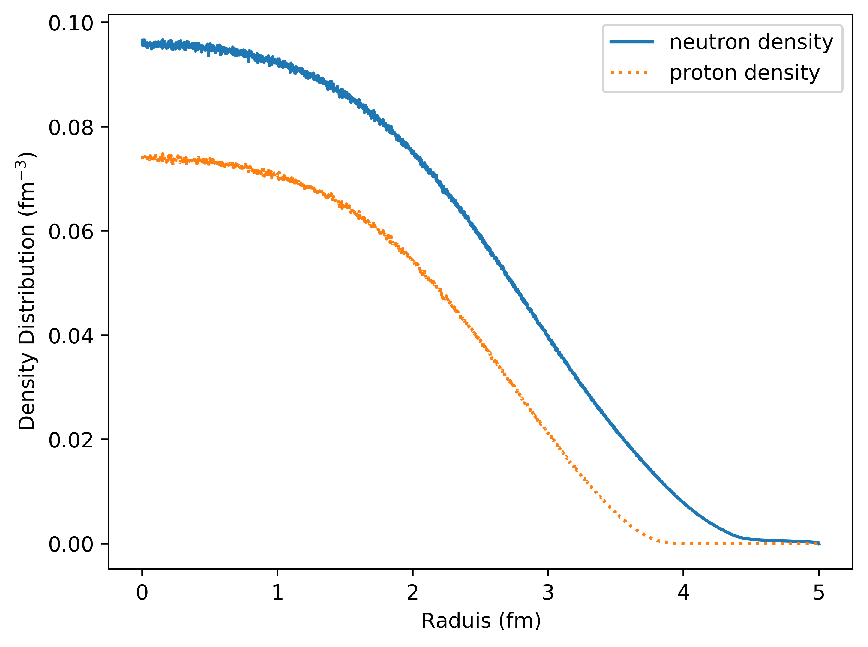}}
	\subfigure[\(^{18}\rm C\)]{\includegraphics[scale=0.6]{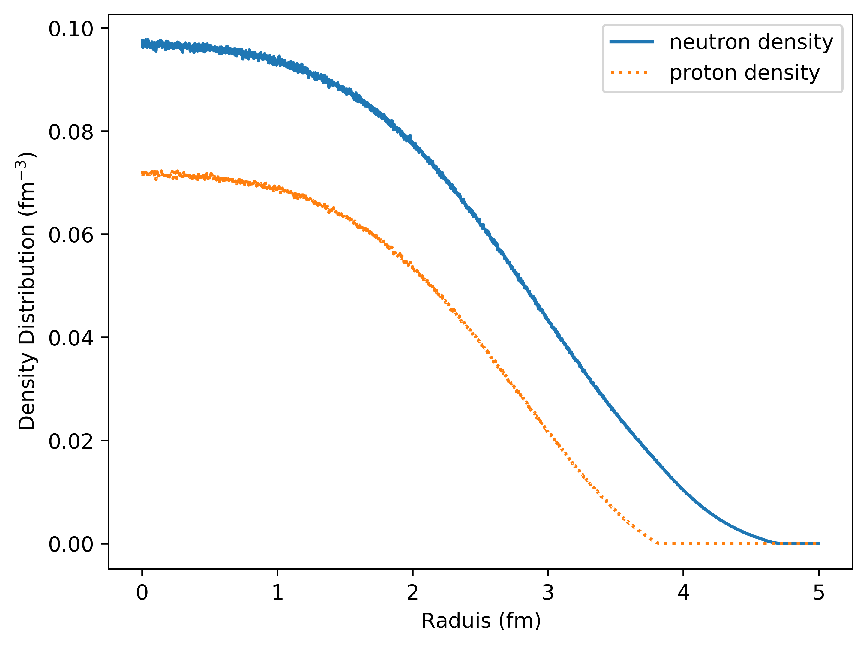}}
	\subfigure[\(^{19}\rm C\)]{\includegraphics[scale=0.6]{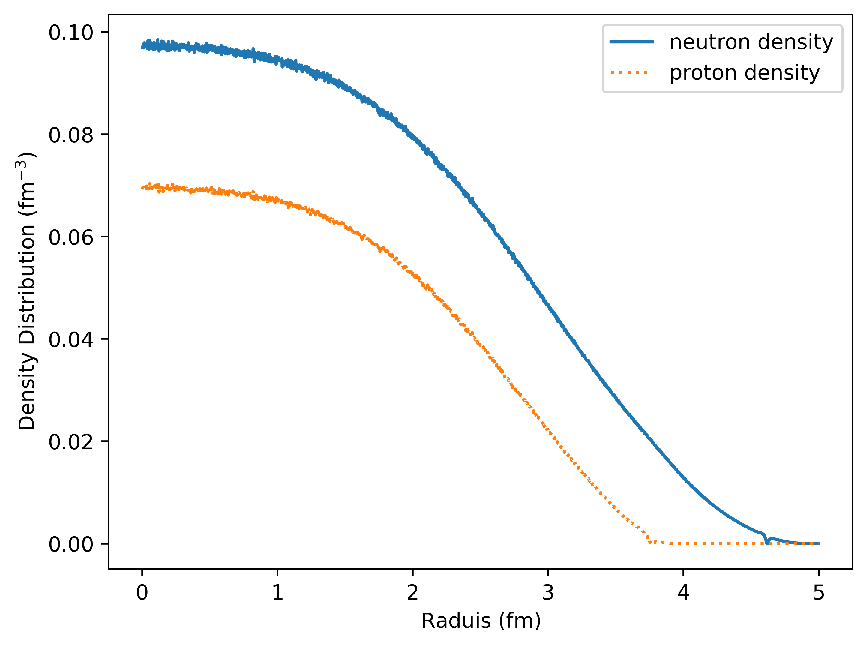}}
	\subfigure[\(^{20}\rm C\)]{\includegraphics[scale=0.6]{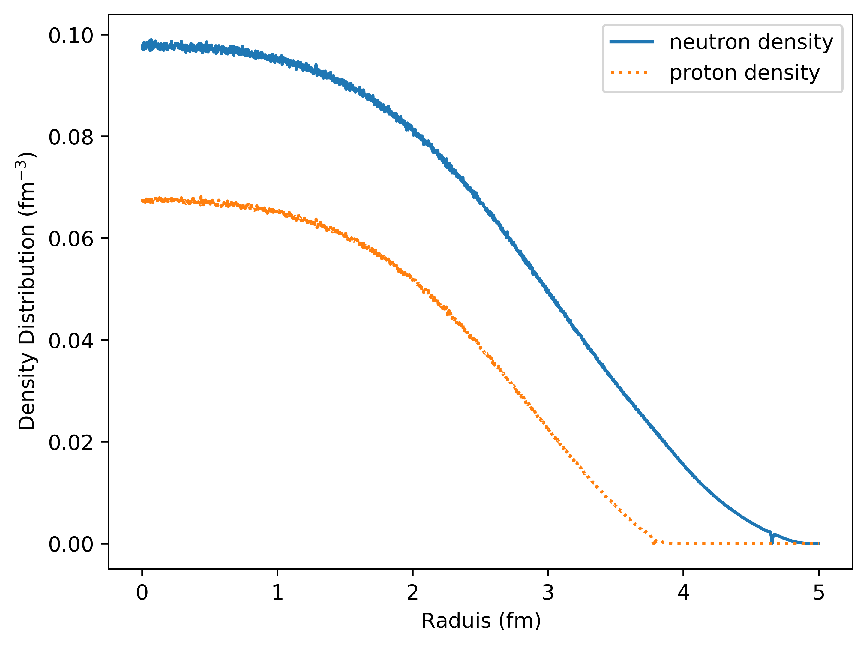}}
	\caption{
		The nucleon density distributions of ground states of different kinds of nucleus.
	}
	\label{fig:nuclei}
\end{figure}
The slight difference between distributions of neutrons and protons in \(^{12}{\rm C}\) is due to EM interactions between protons.

\begin{figure}[htbp]
    \centering
    \subfigure[The pressure dependence on atom-like structure densities.]{\includegraphics[scale=0.6]{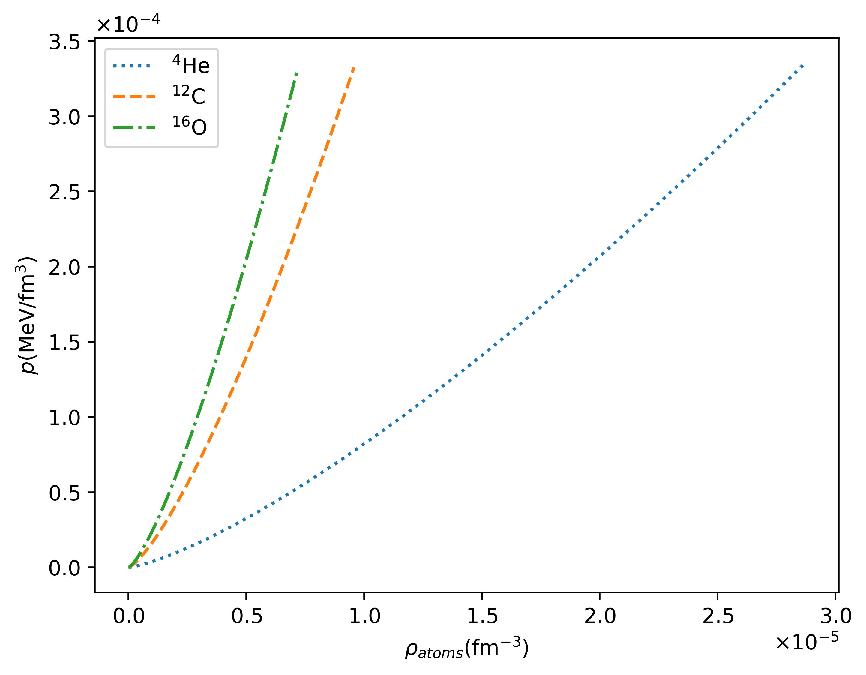}}
    \subfigure[The energy density-pressure relations]{\includegraphics[scale=0.6]{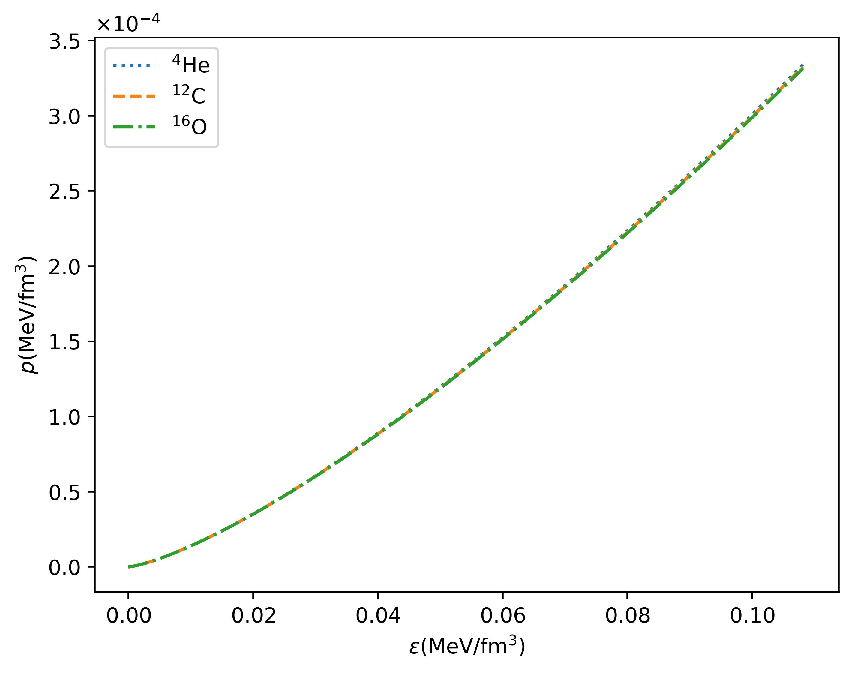}}
    \caption{
        The EOS of WDs composed of different elements.
    }
    \label{fig:EOS}
\end{figure}

\section{The post-Newtonian expansion up to the 2.5PN order}
\label{sec:appPN}

The GW in the frequency-domain ($f$) can be obtained via~\cite{Husa:2015iqa,Isoyama:2020lls},
\be
\tilde{h}(f) & = & \int h(t) e^{i 2 \pi f t} \mathrm{~d} t=A(f) e^{i\left(\psi_{\operatorname{SPA}}(f)-\pi / 4\right)},
\ee
where $\psi_{\mathrm{SPA}}(f)=2 \pi f t(f)-\psi(f)$. $\psi(f)$ can be decomposed into point and tidal parts as $\psi(f) = \psi_{\mathrm{pp}}(f)+ \psi_{\text {tidal }}(f)$. The same treatment applies to amplitude $A(f)=A_{\mathrm{pp}}(f)+A_{\text {tidal }}(f)$. 
The explicit expressions for the amplitude and phase are~\cite{Isoyama:2020lls,Buonanno:2009zt,Kawaguchi:2018gvj}
\be
A_{\rm pp}(f) & \simeq & \frac{\mathcal{M}^{5 / 6}}{r} \sqrt{\frac{2}{3 \pi^{1 / 3}}} f^{-7 / 6}\left(1+O\left(f^{2 / 3}\right)\right)\ ,
\ee
where \(\mathcal{M}=M \nu^{3 / 5}\) is the chirp mass with \(M=m_1+m_2\), $\nu=m_1 m_2 / M^2$, and \(r\) being the observation distance. Meanwhile,
\be
A_{\text {tidal }}(f)=\sqrt{\frac{5 \pi \nu}{24}} \frac{M^2}{r} \tilde{\Lambda} v^{-7 / 2}\left(-\frac{27}{16} v^{10}-\frac{449}{64} v^{12}-4251 v^{15.780}\right),
\ee
where $\tilde{\Lambda}=\frac{16}{13} \frac{\left(m_1+12 m_2\right) m_1^4 \Lambda_1+\left(m_2+12 m_1\right) m_2^4 \Lambda_2}{M^5}$ is the tidal formation of inspirals, $v=(\pi M f)^{1 / 3}$.
Moreover,
\be
\psi_{\mathrm{pp}}(f) & = & 2 \pi f t_c-\phi_c-\frac{\pi}{4}+\frac{3}{128 \nu v^5}\left[1+\frac{20}{9}\left(\frac{743}{336}+\frac{11}{4} \nu\right) v^2-16 \pi v^3 \right.\nonumber\\
& &\left. {}\qquad\qquad\;\;\;\;\;\;\;\;\;\;\;\; +10\left(\frac{3058673}{1016064}+\frac{5429}{1008} \nu+\frac{617}{144} \nu^2\right) v^4+\pi\left(\frac{38645}{756}-\frac{65}{9} \nu\right)\left\{1+3 \log \left(\frac{v}{v_{\text {lso }}}\right)\right\} v^5\right]\ ,
\ee
where \(t_c\) and \(\phi_c\) are chosen to be zero, meanwhile \(v_{\text {lso }}=\sqrt{6}\) is the last-stable-orbit termination condition defined by the Schwarzschild metric, and

\begin{equation}
\psi_{\text {tidal }}(f)=\frac{3 v^5}{128 \nu}\left[-\frac{39}{2} \tilde{\Lambda}\left(1+12.55 \tilde{\Lambda}^{2 / 3} v^{8.480}\right)\right] \times\left(1+\frac{3115}{1248} v^2-\pi v^3+\frac{28024205}{3302208} v^4-\frac{4283}{1092} \pi v^5\right)\ .
\end{equation}
\end{widetext}

\bibliography{bwd}

\end{document}